\documentclass[apj]{emulateapj}

\usepackage{apjfonts}
\usepackage{epsfig}
\usepackage{graphicx}
\usepackage{url}
\usepackage{natbib}
\usepackage{float}
\usepackage{amsmath}

\begin{document}

\shorttitle{Distances to Stars With LAMOST Stellar Parameters}
\shortauthors{Carlin et al.}

\title{Estimation of distances to stars with stellar parameters from LAMOST}

\author{
Jeffrey L. Carlin\altaffilmark{1,2}, 
Chao Liu\altaffilmark{3},
Heidi Jo Newberg\altaffilmark{1},
Timothy C. Beers\altaffilmark{4},
Li Chen\altaffilmark{5},
Licai Deng\altaffilmark{3},  
Puragra Guhathakurta\altaffilmark{6},
Jinliang Hou\altaffilmark{5},
Yonghui Hou\altaffilmark{7}
S\'ebastien L\'epine\altaffilmark{8},
Guangwei Li\altaffilmark{3}
A-Li Luo\altaffilmark{3},
Martin C. Smith\altaffilmark{5},
Yue Wu\altaffilmark{3},
Ming Yang\altaffilmark{3},
Brian Yanny\altaffilmark{9},
Haotong Zhang\altaffilmark{3},
Zheng Zheng\altaffilmark{10}
}

\altaffiltext{1}{Department of Physics, Applied Physics and Astronomy, Rensselaer Polytechnic Institute, Troy, NY 12180, USA, jeffreylcarlin@gmail.com}
\altaffiltext{2}{Department of Physics and Astronomy, Earlham College, Richmond, IN 47374, USA}
\altaffiltext{3}{Key Lab of Optical Astronomy, National Astronomical Observatories, Chinese Academy of Sciences, Beijing 100012, China}
\altaffiltext{4}{Department of Physics and JINA: Joint Institute for Nuclear Astrophysics, University of Notre Dame, 225 Nieuwland Science Hall, Notre Dame, IN 46556, USA}
\altaffiltext{5}{Shanghai Astronomical Observatory, 80 Nandan Road, Shanghai 200030, China}
\altaffiltext{6}{UCO/Lick Observatory, Department of Astronomy and Astrophysics, University of California, Santa Cruz, CA 95064, USA}
\altaffiltext{7}{Nanjing Institute of Astronomical Optics \& Technology, National Astronomical Observatories, Chinese Academy of Sciences, Nanjing 210042, China}
\altaffiltext{8}{Department of Physics and Astronomy, Georgia State University, 25 Park Place, Suite 605, Atlanta, GA 30303, USA}
\altaffiltext{9}{Fermi National Accelerator Laboratory, P.O. Box 500, Batavia, IL 60510, USA}
\altaffiltext{10}{Department of Physics and Astronomy, University of Utah, UT 84112, USA}

\begin{abstract}
We present a method to estimate distances to stars with spectroscopically derived stellar parameters. The technique is a Bayesian approach with likelihood estimated via comparison of measured parameters to a grid of stellar isochrones, and returns a posterior probability density function for each star's absolute magnitude. This technique is tailored specifically to data from the Large Sky Area Multi-object Fiber Spectroscopic Telescope (LAMOST) survey. Because LAMOST obtains roughly 3000 stellar spectra simultaneously within each $\sim5^\circ$-diameter ``plate'' that is observed, we can use the stellar parameters of the observed stars to account for the stellar luminosity function and target selection effects. This removes biasing assumptions about the underlying populations, both due to predictions of the luminosity function from stellar evolution modeling, and from Galactic models of stellar populations along each line of sight. Using calibration data of stars with known distances and stellar parameters, we show that our method recovers distances for most stars within $\sim20\%$, but with some systematic overestimation of distances to halo giants. We apply our code to the LAMOST database, and show that the current precision of LAMOST stellar parameters permits measurements of distances with $\sim40\%$ error bars. This precision should improve as the LAMOST data pipelines continue to be refined.
\end{abstract}

\keywords{Galaxy: stellar content, Galaxy: structure, stars: distances, surveys (LAMOST)}

\section{Introduction}

The Large Sky Area Multi-object Fiber Spectroscopic Telescope (LAMOST) survey \citep{czc+12, dnl+12, lzz+12, zzc+12} has thus far obtained medium-resolution ($R \sim 2000$) spectra for over 3~million stars, on its way to a goal of acquiring some 6-8~million stellar spectra over a planned 5-year survey. Such a vast survey provides an invaluable resource for studies of Milky Way stellar dynamics. LAMOST data have been used to explore kinematical asymmetries in the nearby disk \citep{cdn+13}, nearby stellar moving groups (\citealt{zzc+14,xia15}), high-velocity \citep{zcl+14} and hyper-velocity stars \citep{zcb+14}, and stellar kinematics in the solar neighborhood \citep{tlc+15}. More distant halo structure is also accessible via red giant branch (RGB) stars observed by LAMOST. However, RGB stars can be difficult to identify among the much more numerous foreground dwarfs. \citet{ldc+14} developed a method to select K-type giants from their LAMOST spectra, and a technique to identify M giants (which are not processed in the main LAMOST stellar parameters pipeline) has been developed by \citet{zll+15}; see also the study of the Sagittarius tidal stream M giants by Li et al., {\it in preparation}). All of these studies require an estimate of the distances to the stars involved in order to place them within the structure of the Galaxy. To use the LAMOST data to understand the dynamics of the Milky Way, it is thus vitally important to devise a method for determining stellar distances.

One of the main outputs of a survey such as LAMOST is the spectroscopically derived line of sight (radial) velocity (RV) for each star. In order to exploit this velocity information as a probe of Galactic dynamics, one must know the three-dimensional position of each star within the Galaxy. The position on the sky is, of course, well known, but estimating the third dimension of each star's position (its distance) is non-trivial. Furthermore, even if reliable proper motions are known for each star, combining these with the RVs to derive a three-dimensional space motion for each star requires knowing its distance. A survey such as LAMOST inevitably contains numerous nearby (mostly disk) dwarfs, with a smaller fraction of intrinsically bright RGB stars (e.g., K giants; see \citealt{ldc+14}) that can be used to study more distant structures in the halo of the Milky Way. Thus, in order to fully exploit LAMOST data for studies of Galactic structure, we require not only reliable estimates of stellar parameters, but a robust estimate of the distance to each star, based on available photometry and information gleaned from its spectrum. 

For each star whose spectrum has sufficient signal-to-noise (S/N), the LAMOST pipeline (see, e.g., \citealt{wll+11, lzz+12, wld+14}) derives effective temperature ($T_{\rm eff}$), surface gravity ($\log{g}$), and metallicity ([Fe/H]). When these stellar parameters are known for a single star, it is typical to compare the measured parameters to theoretical isochrones, and use the best theoretical match to estimate the absolute magnitude of the star in question. This is fairly straightforward for a single star, but to robustly determine distances by this technique for large numbers of stars in an automated way presents a challenge. 

Recently, a variety of techniques have been presented for deriving distances to large numbers of stars with stellar parameters resulting from spectroscopic surveys such as RAdial Velocity Experiment (RAVE; \citealt{szs+06,kgs+13}), SDSS/SEGUE (\citealt{yrn+09}), and SDSS/APOGEE. 
These include a $\chi^2$ minimization routine that compares RAVE stellar parameters to a grid of isochrones \citep{bsh+10}, and a modification of this technique \citep{zmb+10} to account for the stellar luminosity function of the isochrones. Bayesian methods that include models of Galactic stellar populations as priors (e.g., \citealt{bbs+11,bbk+14}) have also been applied to RAVE data. Distances presented by the SEGUE Stellar Parameters Pipeline (SSPP; \citealt{lbs+08}) are based on empirical fits from \citet{bcy+00} to globular cluster fiducial sequences, and require separate calibrations for stars in different evolutionary stages. An alternate method that has been applied to SEGUE halo K-giants uses a Bayesian approach to account for the luminosity function and metallicity distribution \citep{xmr+14}. Distances for stars observed by APOGEE have thus far been limited to well-characterized stars such as red-clump stars \citep{bnr+14} and red giants in the Kepler field that have asteroseismic surface gravities \citep{rgm+14}.

Our method of deriving distances from LAMOST spectroscopic parameters was chosen to avoid introducing assumptions about stellar populations and their distribution in the Galaxy. We simply want the best empirical estimate of the distance, so that we can {\it use} this to explore the distribution of stellar populations in the Milky Way. The distribution of observed stellar parameters will be biased by the method used to select targets (e.g., the color and magnitude selection criteria), the intrinsic properties (including the stellar parameters and luminosity function) of the Galactic sub-populations sampled in each observed region, and perhaps even the observing conditions under which each spectrum was obtained. One benefit of LAMOST is that each star is observed as part of a ``plate'' on which $\sim3000$ stars are simultaneously targeted. Each plate covers a narrow magnitude range and has a simple target selection function (see, e.g., \citealt{cln+12,lyh+14}), making it possible for us to use the observations themselves to account for both the selection function and the stellar luminosity function in our distance estimates. In practice, the selection function has not remained the same throughout the LAMOST survey, making it impossible to back out the probability of observing a star with given properties explicitly. We account for the effects of target selection by using the empirically measured distribution of stellar parameters on each plate as a prior for the likelihood of finding a star of a given surface gravity at a given color and magnitude. In this way, we are explicitly including the observed $\log{g}$ for all stars on a plate to derive an estimate of our expectations along each line of sight, thus removing selection biases and the effects of differently sampled Milky Way populations along each line of sight.

This paper is outlined as follows. Section~\ref{sec:dist_method} discusses the techniques we have developed for deriving distances to stars with LAMOST stellar parameters. In Section~\ref{sec:verification} we verify the effectiveness of our technique using several catalogs from the literature, as well as simulated data sets. We follow with a brief illustration of the results from applying our algorithms to the entire LAMOST data set of $\sim1.8$~million stellar spectra in Section~\ref{sec:lamost_dist}. Finally, we conclude with some remarks about the utility of this method for Galactic structure science with LAMOST.

\section{Distance determination methods}\label{sec:dist_method}

We derive distances to stars by comparing measured stellar parameters to a grid of synthetic isochrones. Initially, a simple $\chi^2$ method is tested. Though the results from this algorithm seem reasonable, it is difficult to derive reasonable uncertainties. We thus turn to a Bayesian technique, which also has the advantage of allowing the priors to be easily adjusted in the future as desired.

\subsection{Adopted Isochrones}\label{sec:iso}

We began by creating a grid of isochrones from the Dartmouth Stellar Evolution Database \citep{dcj+08}. This particular set of isochrones was chosen in part because it more accurately reproduces the lower main sequence in SDSS colors than other systems \citep{fc12, fjc14}, and in part simply for the convenience with which one can generate a custom grid of isochrones in the Dartmouth system.\footnote{Unlike most available isochrone systems, the Dartmouth isochrones include atomic diffusion (see \citealt{dcj+08} and references therein), and are thus more reliable for age estimates. While measuring stellar ages is not a current concern of ours, we may wish to use the grid for this purpose in the future.} We tested our programs with Padova isochrones \citep{gbb+02}, and found that the differences in derived distances are less than a few percent, with most of the discrepancies at the cooler end of the main sequence. Our adopted grid contains isochrones ranging from $-2.5 <$ [Fe/H] $< +0.5$ in 0.1~dex increments, and 1-15 Gyr in linearly spaced 1~Gyr increments. 
All isochrones were generated with $[\alpha$/Fe] = 0.0.
Isochrone grids were generated for the SDSS $ugriz$ and the $UBVRIJHK_sK_p$ photometric bands, in order to allow for the use of a variety of input magnitudes to derive distances. We removed low-mass stars ($M < 0.4 M_\odot$) and all evolutionary stages other than main sequence, subgiant, and RGB. Other evolutionary stages (e.g., horizontal branch) are not well classified at present by LAMOST spectra, and are also not well represented in the isochrones, so we chose to excise them and keep only ``normal,'' well-behaved stars.

Before using the grid of isochrones for derivation of distances, it was interpolated onto a regularly spaced distribution in absolute magnitude. Because we need an absolute magnitude that corresponds to each metallicity, age, surface gravity, and temperature in the grid, we begin by creating a dummy array of absolute magnitudes spanning the relevant range ($8 > M_{\rm K_S} > -8$ for 2MASS colors/magnitudes, and $12 > M_r > -4$ for SDSS), in increments of 0.02 magnitudes. For each combination of the 15 age steps and 31 steps in [Fe/H] making up our grid, we then use a cubic spline interpolation to map the effective temperature ($T_{\rm eff}$) and surface gravity ($\log{g}$) behavior as a function of absolute magnitude. In this way, we create a grid with identical values of absolute magnitude for each age/metallicity combination, which then simply reflects the $T_{\rm eff}$ and $\log{g}$ of a theoretical star at that age/metallicity that would have each value of absolute magnitude in the array. In other words, at each absolute magnitude, we create arrays with all combinations of age, [Fe/H], temperature, and surface gravity that are predicted by the isochrone grid.

\subsection{Chi-squared Technique}

The goal is to take the measured stellar parameters $T_{\rm eff}$, $\log{g}$, and [Fe/H], along with known photometric magnitudes and colors, and derive a distance to each star. We employ near-infrared 2MASS \citep{scs+06} magnitudes and colors from here onward, but these can be replaced with magnitudes from any other system (e.g., SDSS) for which the Dartmouth isochrones have been calculated. We chose 2MASS because $\sim97\%$ of the objects in the LAMOST catalog have matches in 2MASS, thus providing a uniform input catalog (note that the majority of the objects that do not have 2MASS counterparts are at the faintest magnitudes reached by LAMOST). The use of 2MASS also simplifies comparisons to other catalogs that may not overlap the SDSS footprint or magnitude range. 

Assuming we have measured input parameters (``observables'') $T_{\rm eff}$, $\log{g}$, [Fe/H], and $(J-K_S)$ (or any other input color), and associated errors $\sigma_{T_{\rm eff}}, \sigma_{\rm{log}(g)}, \sigma_{\rm [Fe/H]}$, and $\sigma_{(J-K_S)}$, we can define a $\chi^2$ statistic:

\begin{equation}
\chi^2 = \sum\limits_{i=1}^n \frac{(O_i - O_{i, \rm{mod}})^2}{\sigma_{O_i}^2},
\end{equation}

\noindent where the $O_i$ are the observables ($n=4$ in this example) with associated errors $\sigma_{O_i}$, and $O_{i, \rm{mod}}$ are the isochrone model parameters corresponding to each observable. To determine the best value for our data, we find the model point at which $\chi^2$ is a minimum. The distance modulus then simply consists of the difference between the model star's absolute magnitude at this best-fit point and the input (measured) magnitude.

To account for errors on the observables, while also deriving an uncertainty on our derived distance, we resample $N$ times (we adopted $N$=100 per star after confirming that this relatively small number of samples produces nearly identical results as $N=1000$ per star, while keeping computation times reasonable) from within Gaussian-distributed errors on each of the parameters, then minimize $\chi^2$ for each sample. The mean and standard deviation of the probability distribution function (PDF) for the absolute $K_S$ magnitude ($M_K$) from this Monte Carlo process are measured for each star. We then combine these with the observed $K_S$ magnitude to derive the distance and its error. At this point, the uncertainties on many of the distances were found to be unrealistically high (often greater than 100\% for stars from LAMOST); this is likely because the errors on the stellar parameters quoted in the LAMOST catalog are overestimated (as found by \citealt{lbc+15} via comparison to SDSS spectra of stars in common between the surveys). Although the grid spacing of 0.1~dex in [Fe/H] is half the smallest metallicity uncertainty we have considered ($\sigma_{\rm [Fe/H]} = 0.2$), it may also be that finer grid spacing (in both age and metallicity) would reduce the scatter in Monte Carlo-resampled distance estimates. Furthermore, though we tested resampling $N=100$ and 1000 times and found little effect on the derived distances and their errors, it is also possible that even larger samples are required to fully reproduce the correct PDF of $M_K$ for each star; this was not explored further because of the computationally prohibitive cost of resampling 10,000 or more times per star.

\subsection{Bayesian Technique}

In order to obtain a posterior PDF for the distance rather than a simple distance estimate with associated error bar, we adopt a Bayesian method. This more readily allows for statistical studies of Galactic populations that require the full PDF. We choose to keep the priors in our method extremely simple, unlike methods (e.g., \citealt{bbs+11,bbk+14}) that consider priors based on models of Milky Way stellar populations. Because we intend to use the derived distances to study the kinematics and density distributions of Milky Way stellar populations and their metallicity distribution functions, we wish to avoid priors that assume any uncertain properties of these populations. It is simple to incorporate more complex priors into our algorithm in the future, should we wish to do so.

Consider a vector of observed stellar parameters ${\bf O} = (T_{\rm eff}, \log{g}, {\rm [Fe/H]}) \equiv (T, G, Z)$ with measurement errors ${\bf \sigma_O}$. Assume that these can be mapped via stellar models onto a vector of intrinsic properties ${\bf X} =$ (age, mass, metallicity)$ \equiv (A, M_0, Z_0)$ that together determine the evolution of each star.\footnote{The Dartmouth isochrones are given explicitly in [Fe/H] rather than $Z$. However, the values of $Z$ and $Y$ (the helium fraction; calculated as $Y = 0.245 + 1.54 Z$) are given in the header of each isochrone, with details about the initial compositions of the models given in \citet{dcj+08}. Thus, if desired, one can use the standard definition of [Fe/H]$=\log{(Z/X)/(Z/X)_\odot}$, with standard solar values from \citet{gs98} of $(Z/X)_\odot = 0.0267$ (initial) and $(Z/X)_\odot = 0.0229$ at the current solar age, to find $Z$.}
These intrinsic properties combined with the stellar models give the absolute magnitude distribution $M_{\rm abs} (A, M_0, Z_0)$. The mapping from observables ${\bf O}$ to $M_{\rm abs}$ represents a convolution of the intrinsic luminosity function and the ways in which the selection function has sampled this luminosity function. We will denote the selection function as $S$.  Stellar models relate $A$ to ${\bf O}$; we do not include any explicit dependence of model parameters on $A$ because we do not know anything about the ages of observed stars in advance (i.e., we take a uniform prior $p (A) = 1$). 
Given a set of stellar models, observed stellar parameters, and the selection function, a PDF for absolute magnitude can be derived. The full PDf is:

\begin{equation}
p(M_{\rm abs}, {\bf O}, {\bf \sigma_O}, S) = p(M_{\rm abs} | {\bf O}, {\bf \sigma_O}, S) p({\bf O}, {\bf \sigma_O}, S)
\end{equation}

\noindent This can be rewritten as:

\begin{equation}
\begin{aligned}
p(M_{\rm abs}, {\bf O}, {\bf \sigma_O}, S) &= p(M_{\rm abs} | {\bf O}, {\bf \sigma_O}, S) p({\bf O}, {\bf \sigma_O}, S) \\
& = p({\bf O} | M_{\rm abs}, {\bf \sigma_O}) p({\bf \sigma_O} | M_{\rm abs}) p(M_{\rm abs} | S) p(S),
\end{aligned}
\end{equation}

\noindent where in the first term on the right-hand side, we have removed the dependence on $S$, because once the star has been observed, the likelihood $p({\bf O}| M_{\rm abs}, {\bf \sigma_O})$ no longer depends on the selection function. Rearranging, we obtain:

\begin{equation}
p(M_{\rm abs} | {\bf O}, {\bf \sigma_O}, S) = \frac{  p({\bf O} | M_{\rm abs}, {\bf \sigma_O}) p({\bf \sigma_O} | M_{\rm abs}) p(M_{\rm abs} | S) p(S) }{p({\bf O}, {\bf \sigma_O}, S)}
\end{equation}

We assume that the measured errors ${\bf \sigma_O}$ are independent of $M_{\rm abs}$. This may not strictly be true -- the errors may indirectly depend on intrinsic properties of the star (e.g., low-metallicity stars may have larger $\sigma_Z$, or giants near the RGB tip may have higher $\sigma_G$), which in turn affect the $M_{\rm abs}$. However, this should be a minimal effect for our purposes, so we take $p({\bf \sigma_O} | M_{\rm abs})$ to be independent of $M_{\rm abs}$ (i.e., this term becomes $p({\bf \sigma_O})$). We also neglect the denominator, $p({\bf O}, {\bf \sigma_O}, S)$, which contributes only a normalization factor.
The term $p(M_{\rm abs} | S)$ is the absolute magnitude distribution given the selection function. If there is no selection function, this term would be the luminosity function. This leaves us with a final expression for the posterior PDF of $M_{\rm abs}$:

\begin{equation}
p(M_{\rm abs} | {\bf O}, {\bf \sigma_O}, S) \propto p({\bf O} | M_{\rm abs}, {\bf \sigma_O}) p(M_{\rm abs} | S) p({\bf \sigma_O}) p(S)
\label{pdf_eqn_final}
\end{equation}

We take the likelihood $p({\bf O} | M_{\rm abs}, {\bf \sigma_O})$ to be Gaussian in each of $T, G$, and $Z$, i.e.,

\begin{equation}
p({\bf O} | M_{\rm abs}, {\bf \sigma_O}) \propto \prod_{i=1}^{3}{ \exp { [ -(O_i-O_{\rm mod})^2 / 2 \sigma_{O,i}^2 ]}},
\end{equation}

\noindent where $i=1-3$ to indicate the inclusion of $T, G,$ and $Z$ in the product, and $O_{\rm mod}$ indicates the corresponding parameters for the model isochrone grid (and thus implicitly $M_{\rm abs}$). In practice, this is accomplished by calculating this Gaussian residual for the input star relative to every point in the model isochrone grid. Because every point on the grid with its parameters $O_{\rm mod}$ has an associated $M_{\rm abs}$, the likelihood given by this product of Gaussians can be mapped to a likelihood distribution in $M_{\rm abs}$.

\begin{figure}[!t]
\begin{center}
\includegraphics[width=0.975\columnwidth]{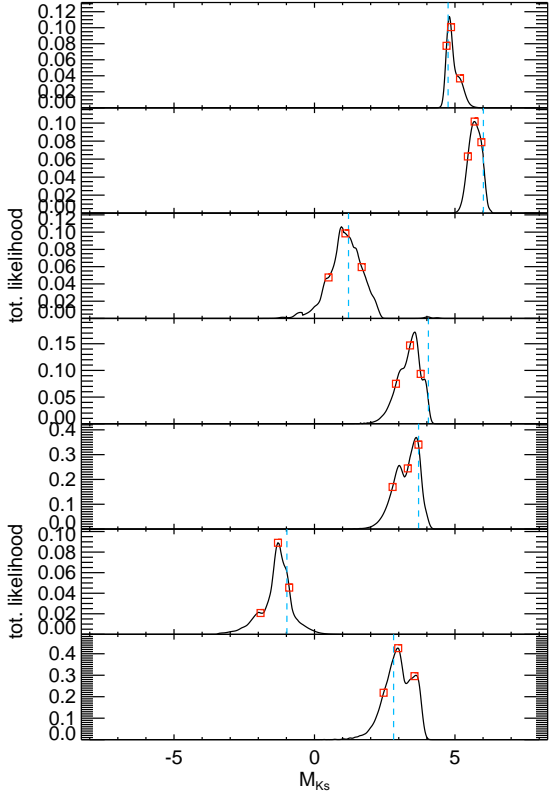}
\caption{Examples of the posterior PDF values from the Bayesian distance code. Each panel shows the likelihood distribution derived from the stellar parameters for a randomly chosen star from the Besan\c{c}on \citep{rrd+03} model catalog. Red squares represent the 15\%, 50\%, and 85\% values in the cumulative PDF, which we take as the best estimate (50\%) and error bars for the absolute magnitude. Dashed blue lines in each panel show the ``true'' absolute magnitude given by the Besan\c{c}on model.  }
\label{samp_likelihood}
\end{center}
\end{figure}

The term $p(M_{\rm abs} | S)$ in Equation~\ref{pdf_eqn_final} encodes the relative numbers of stars as a function of $M_{\rm abs}$, given the selection function $S$ near each star's line of sight; i.e., how likely a star of a certain $M_{\rm abs}$ is to have made it into the sample given the catalog of observed stars.
This term is necessary because the entire luminosity function is not sampled by a given region of color-magnitude space. To account for this selection effect (which depends on color and magnitude, but more importantly on position in the sky), we derive an empirical correction based on the stars actually observed in a given LAMOST plate. To do so, we select stars from the same LAMOST plate that are within 0.25 magnitudes in color and magnitude (e.g., $K_{S,0}$ and $(J-K_S)_0$, or whatever color-magnitude system is being used to derive distances) of the star of interest. For each nearby star, we generate a Gaussian centered at its measured $\log{g}$ with width equal to its associated error, $\sigma_{\log{g}}$, and then normalize its sum to unity to create a PDF. We create a generalized histogram by summing these PDFs for all of the color-magnitude selected stars in the plate, then normalize the resulting $\log{g}$ distribution to yield the probability of finding a star at a given $\log{g}$ value in the vicinity (in both position and color-magnitude) of the star of interest. 
This $\log{g}$ distribution is then mapped onto each input isochrone via interpolation of the $\log{g}-M_{\rm abs}$ relation of the isochrone itself. The resulting histogram in absolute magnitude is normalized to provide the probability of finding stars of a given $M_{\rm abs}$ based on the measured $\log{g}$ distribution. We incorporate this probability distribution as $p(M_{\rm abs} | S)$ in Equation~\ref{pdf_eqn_final} to properly account for the underlying luminosity function along each line of sight, and the selection effects of the survey (which corrects for both the fraction of stars that were selected as a function of color and magnitude as well as the volume of the Galaxy sampled by the selected stars). 

\begin{figure*}[!t]
\includegraphics[width=0.45\textwidth]{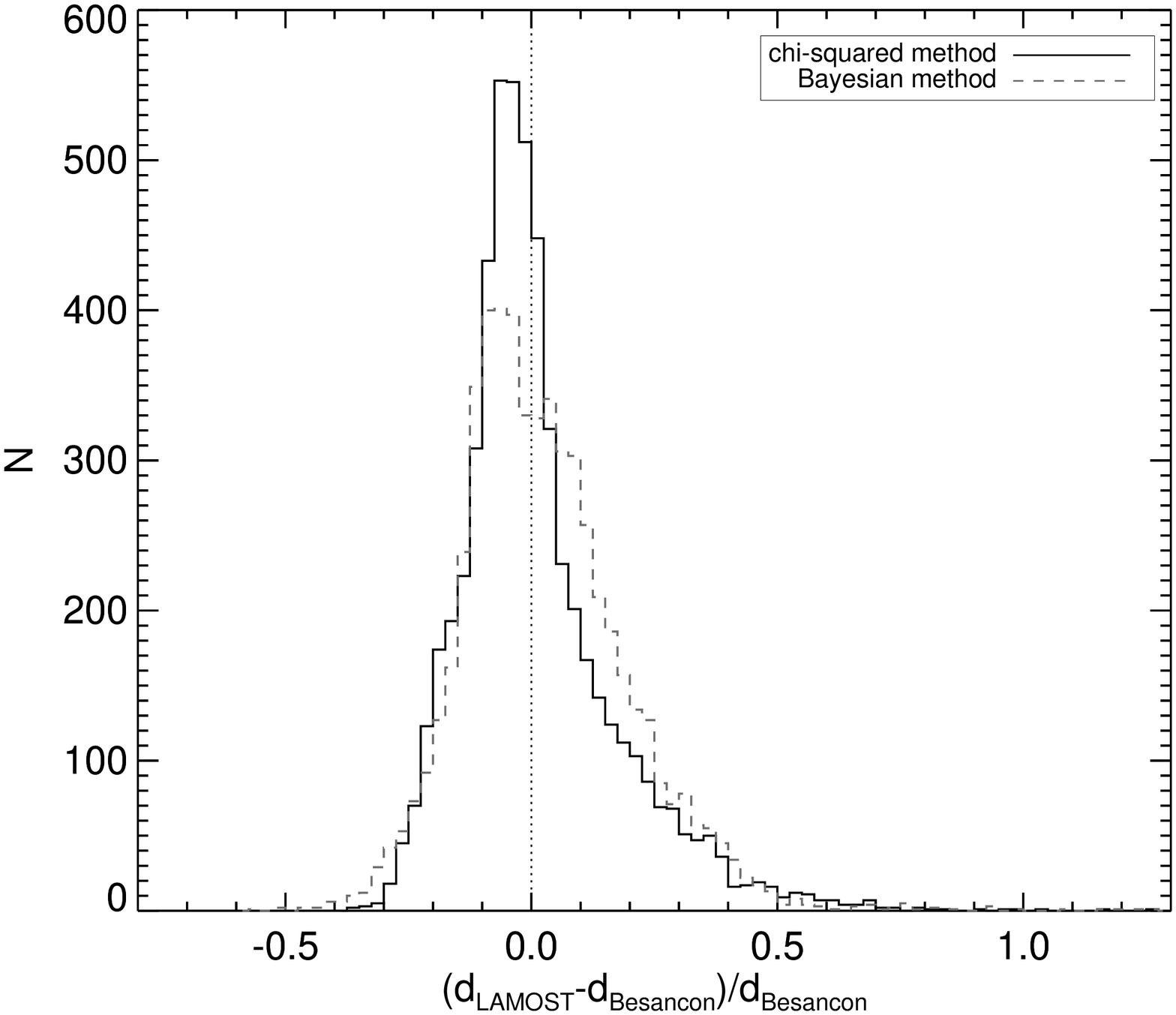}
\includegraphics[width=0.46\textwidth, trim=0.0in 0.5in 0.0in 0.0in]{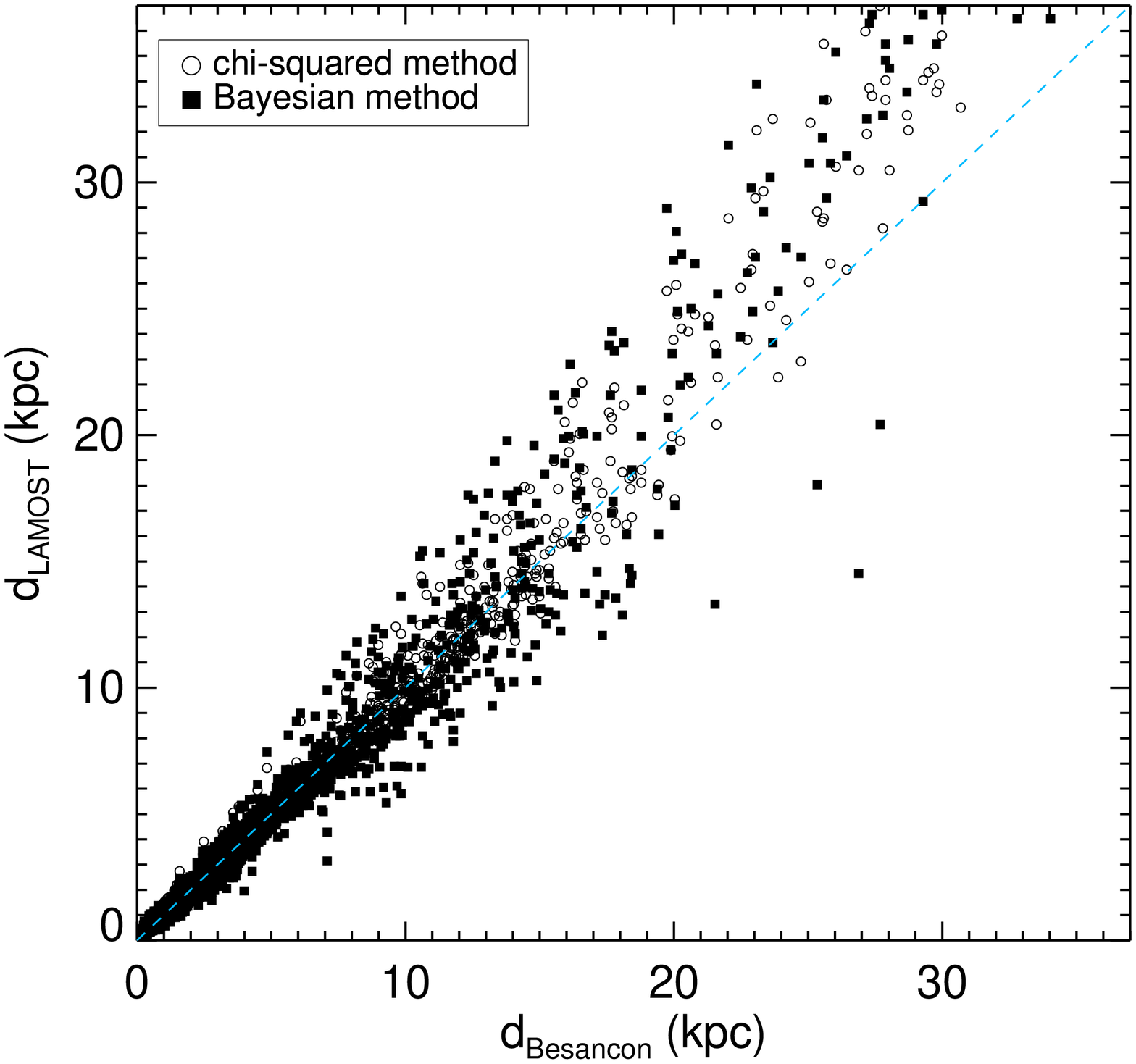}
\caption{Comparison of distances derived by our two algorithms for Besan\c{c}on model galaxy stars in a simulated field at $(l, b) = (180^\circ, 60^\circ)$. The left panel shows the residuals from comparison of the Besan\c{c}on model distances, $d_{\rm Besancon}$, to distances derived from the model stellar parameters using our code, $d_{\rm LAMOST}$. The solid black line is the output from the $\chi^2$ algorithm, and the dashed gray histogram is from the Bayesian code. The results from the Bayesian method have slightly larger scatter than those from the $\chi^2$ algorithm, but little systematic offset is seen in either method. The right panel directly compares the input (Besan\c{c}on) distances to those recovered by the code ($d_{\rm LAMOST}$). Open symbols are $\chi^2$ distances, and filled symbols are results from the Bayesian code. The agreement is good out to $d \sim 15$~kpc. Beyond this distance, our algorithms systematically overestimate the distances by $\sim20\%$ for the majority of stars. We believe this discrepancy is due to our use of isochrones with solar $[\alpha/$Fe]; because metal-poor halo giants are typically alpha-enhanced relative to Milky Way disk stars, a more appropriate alpha-enhanced isochrone grid may remedy this deficiency for halo stars (see Section~4.2 and Figure~\ref{alpha_enhanced_dist} for further exploration of this effect).
}
\label{compare_methods}
\end{figure*}

In the absence of a selection function, $p(M_{\rm abs} | S)$ could be represented by the theoretical luminosity function of the isochrones. For general usage, we include this feature in the code for instances where the parameters of nearby stars are not known.
In the Dartmouth isochrones, the density of points along each isochrone as given encodes equal steps in ``equivalent evolutionary phase'' \citep{bbc+90}. Assuming that this roughly mimics the luminosity function, we calculate the normalized density of points as a function of absolute magnitude for each isochrone in histogram bins of 0.2 magnitudes, and use this as the starting estimate for $p (M_{\rm abs} | S)$ when there are not sufficient nearby, simultaneously observed stars to use for the selection function correction. In tests of the effect of the selection
function correction using 20,000 stars, we found that the fractional change in distance between measurements with/without this correction was less than 10\% for 93\% of the stars, and less than 20\% for 97\% of the stars. However, a small number of stars (~1.4\%; mostly red giants) had
their distances change by more than 30\% between these two methods. As expected, then, the correction is important for the much rarer RGB stars than for the ubiquitous nearby dwarfs in the LAMOST database.

The algorithm to evaluate each of the three terms on the right side of Equation~\ref{pdf_eqn_final} produces an array of points corresponding to all of the theoretical stars in the isochrone grid. Each of these points has an associated $M_{\rm abs}$ and a likelihood value. Thus Equation~\ref{pdf_eqn_final} also produces an array of posterior PDFs for a large grid of absolute magnitudes. We sum the PDF for each $M_{\rm abs}$ value to produce a marginalized PDF in $M_{\rm abs}$ for the input star. This is normalized to produce the final PDF $p(M_{\rm abs} | {\bf O}, {\bf \sigma_O}, S)$; some examples are shown in Figure~\ref{samp_likelihood}. We take the median (i.e., 50th percentile) of this PDF as the best estimate for $M_{\rm abs}$, with uncertainties derived using the $M_{\rm abs}$ corresponding to the 15th and 85th percentile values from the cumulative PDF. We also retain the full PDFs so that they can be used instead of single estimates of distances and their errors.

Figure~\ref{compare_methods} shows a comparison of the results from running the two versions ($\chi^2$ and Bayesian) of the code on stars from a mock galaxy field generated by the Besan\c{c}on \citep{rrd+03} model.\footnote{Throughout this work, we use the notation $d_{\rm LAMOST}$ to refer to distances derived by the algorithm discussed in this paper. Note that this does not mean that these distances are derived using LAMOST data, but rather that our LAMOST {\it distance code} is being applied.}  Uncertainties on the stellar parameters for all stars in the mock catalog were set to $\sigma_{\rm Teff} = 100$~K, $\sigma_{\rm log~g} = 0.3$~dex, and $\sigma_{\rm [Fe/H]} = 0.2$~dex. The left panel compares the residuals (in the sense $(d_{\rm LAMOST} - d_{\rm Besancon})/d_{\rm Besancon}$) of measured distances to the input (model) distances. Both methods recover the input distances fairly well, with an asymmetric tail to high (overestimated) residuals. The right panel of Figure~\ref{compare_methods} compares these distance measurements directly. It is clear from this panel that the large positive residuals are mostly for distant, metal-poor giants, whose distances are overestimated by $\sim20\%$. Having now verified that the $\chi^2$ and Bayesian methods produce similar results, we henceforth use only the Bayesian method to obtain the full distance PDF. 

\begin{figure}[!t]
\begin{center}
\includegraphics[width=0.9\columnwidth]{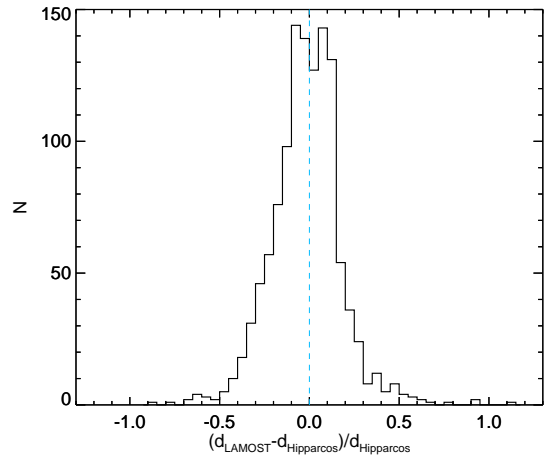}
\caption{Results of running the distance code on 1199 stars from the \citet{gcg+03,gcg+06} catalogs. The histogram shows fractional residuals from comparison of the Hipparcos parallax distances, $d_{Hipparcos}$, to distances derived from the Gray et al. stellar parameters, $d_{\rm LAMOST}$. For reference, the dashed vertical line is one-to-one agreement. Our distances show no systematic offset with respect to the parallax distances, with $\sim17\%$ scatter.}
\label{hiptest}
\end{center}
\end{figure}

\begin{figure*}[!t]
\begin{center}
\includegraphics[width=0.4\textwidth]{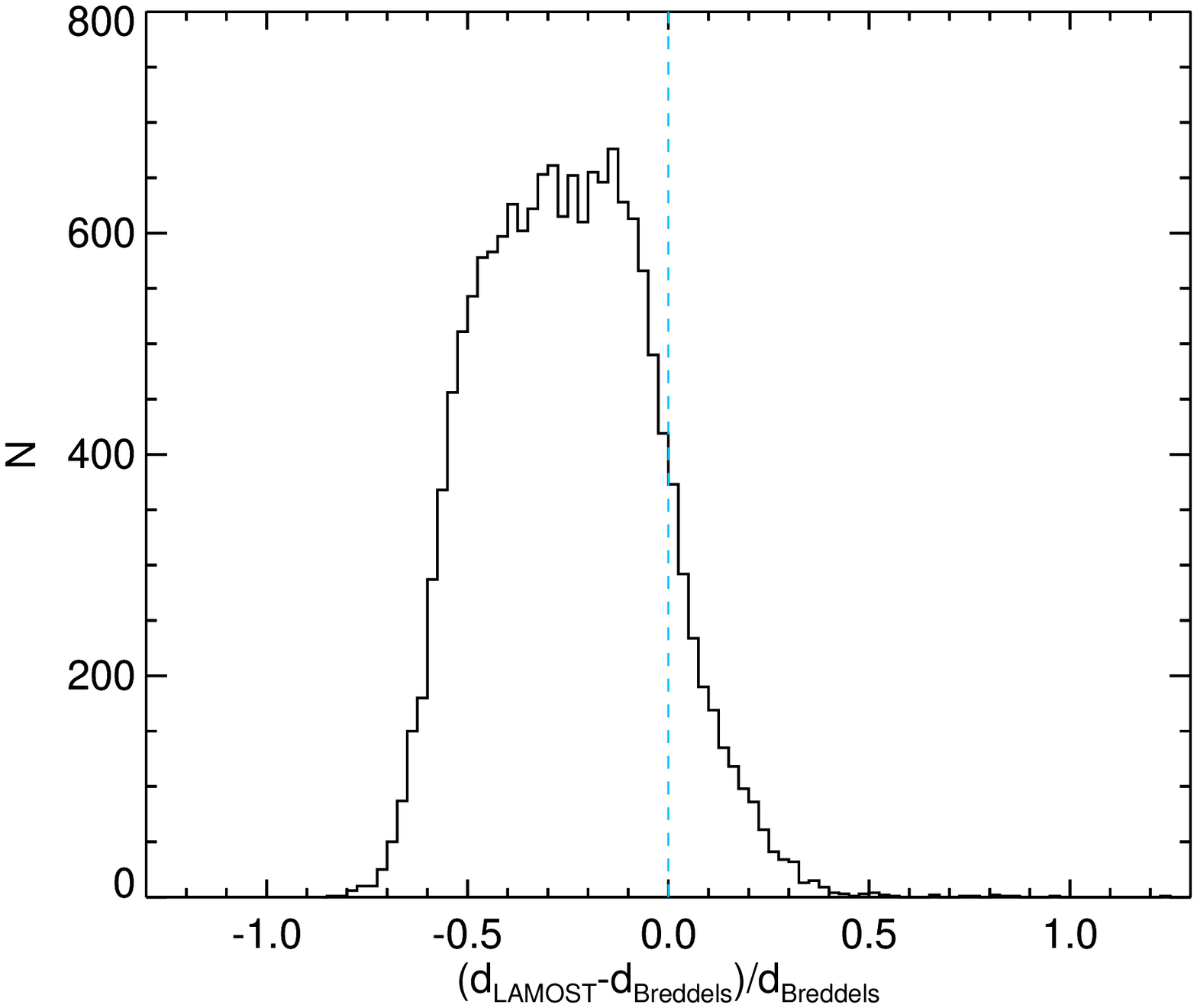}
\includegraphics[width=0.4\textwidth]{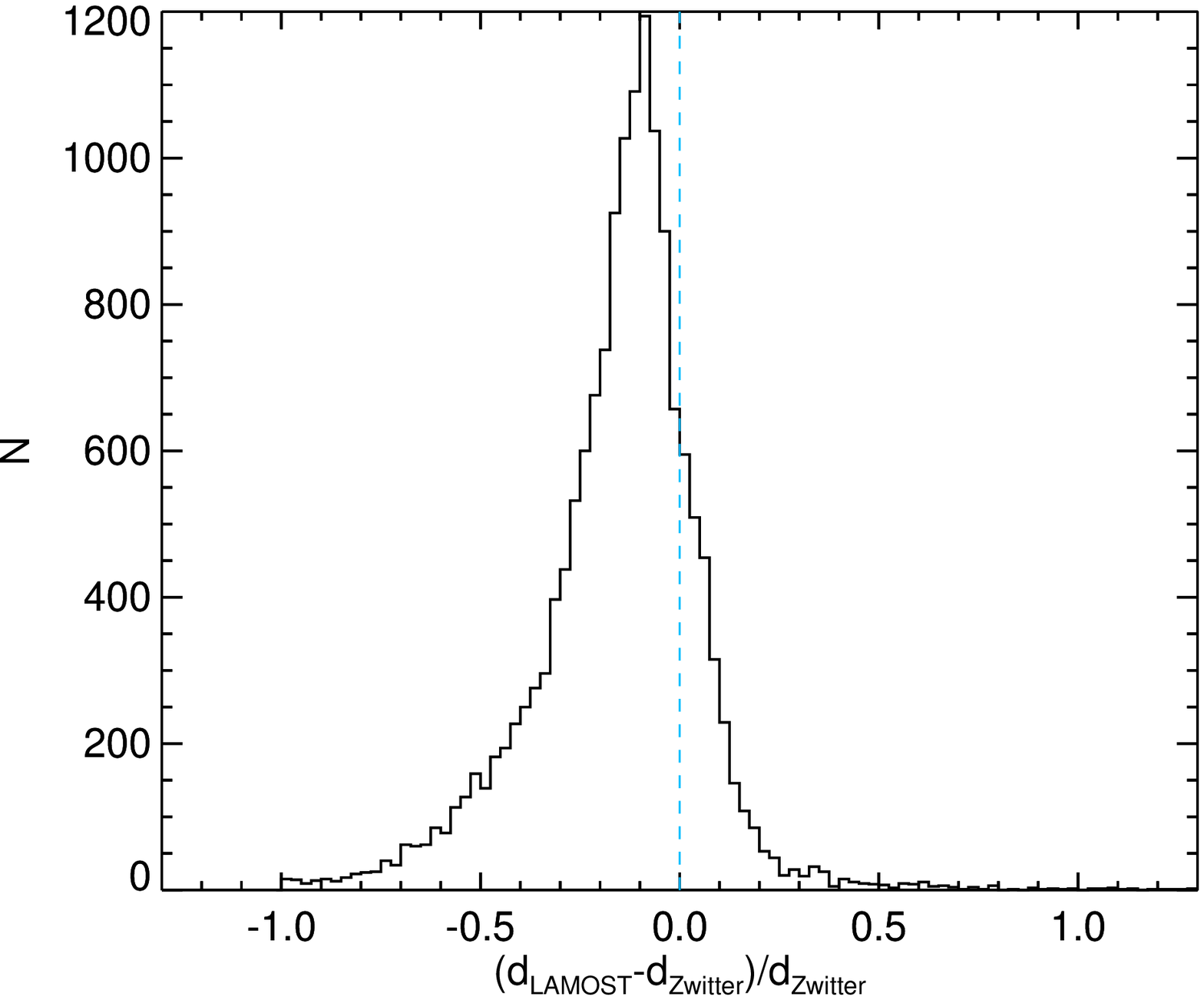}

\caption{Left panel: results of running the distance code on $\sim16,000$ stars from RAVE in the \citet{bsh+10} catalogs. As in Figure~\ref{hiptest}, the panels represent the fractional residuals relative to the catalog values from the Bayesian method. Our distances are systematically shifted by $\sim26\%$ relative to those of \citet{bsh+10}, with $\sim23\%$ scatter.
Right panel: as in the left panel, but for the RAVE sample from \citet{zmb+10}. Our distances are systematically smaller by 
$\sim12\%$, with $\sim16\%$ spread about this value. There are decidedly non-Gaussian tails in these residuals on toward large relative underestimates. 
}
\label{ravetest}
\end{center}
\end{figure*}

\begin{figure*}[!t]
\begin{center}

\includegraphics[width=0.4\textwidth]{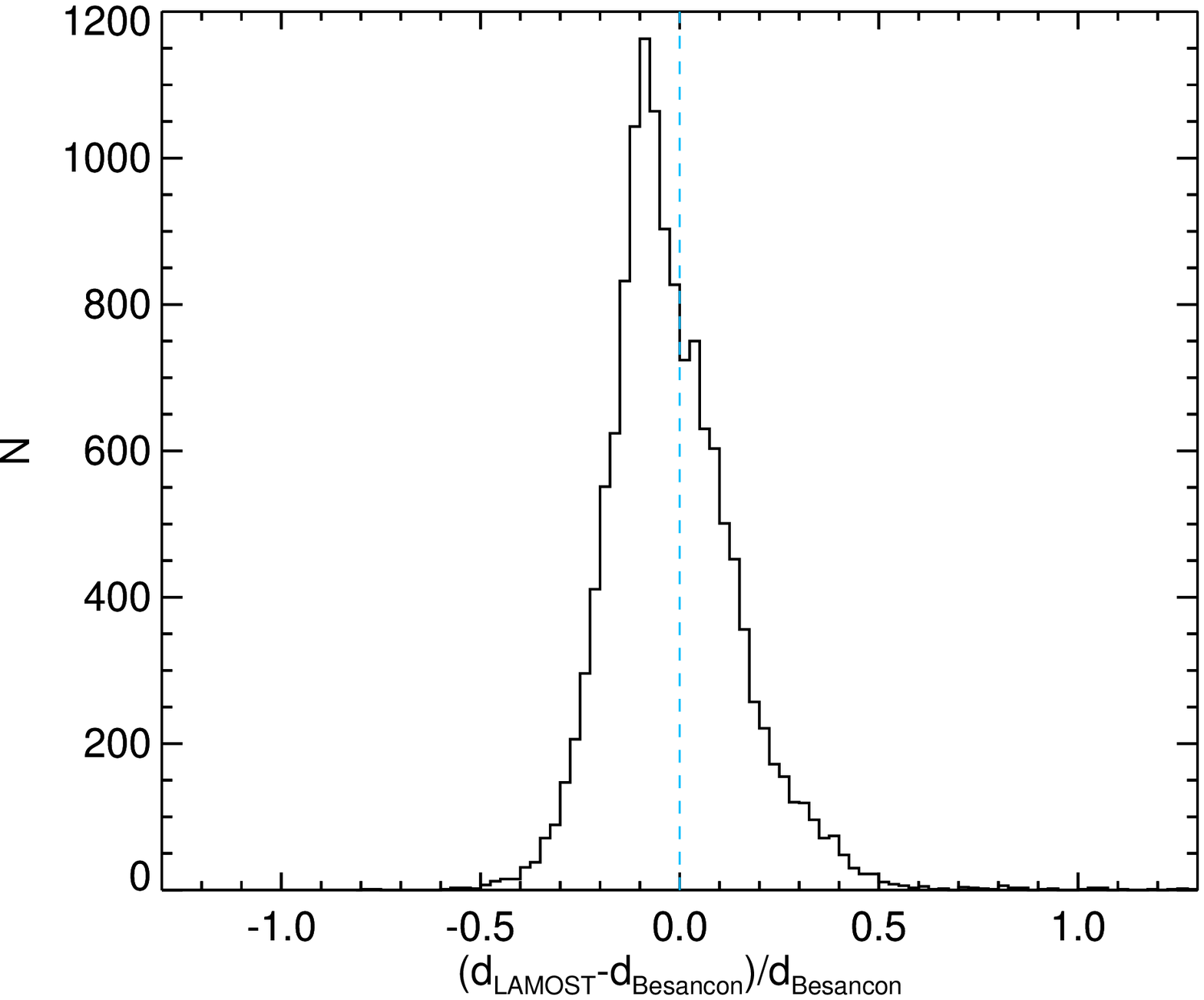}
\includegraphics[width=0.4\textwidth]{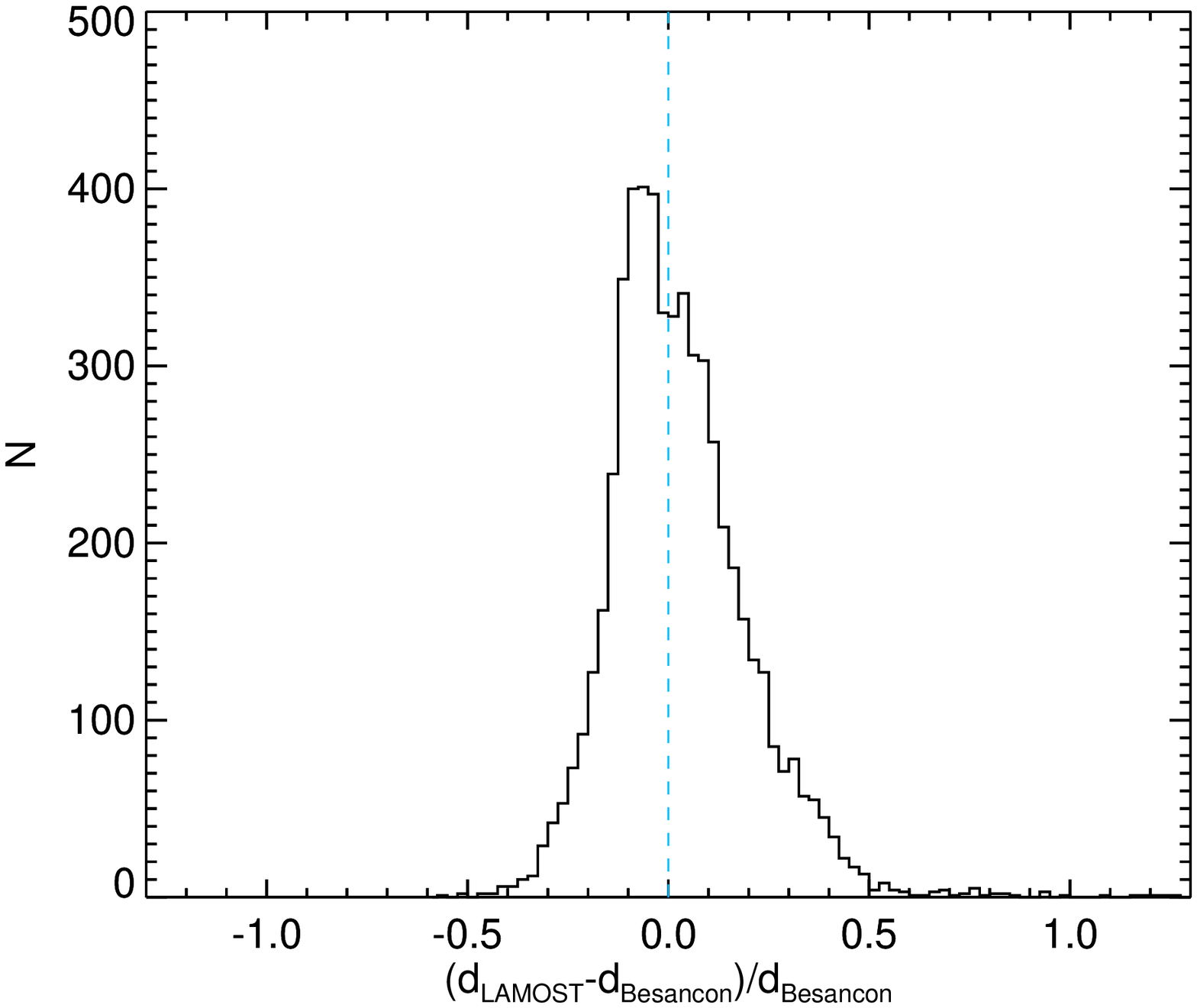}

\caption{Results from running the distance code on two realizations of the Besan\c{c}on Galaxy model. The left panel is for a field at $l=180\arcdeg, b=30\arcdeg$, and the right panel shows a field at $l=180\arcdeg, b=60\arcdeg$. }
\label{disttest}
\end{center}
\end{figure*}

\section{Verification of the methods}\label{sec:verification}

To test the code, we need a sample of stars with known spectroscopic parameters and distances. For this, we use the \citet{gcg+03,gcg+06} measurements of stellar parameters for Hipparcos stars within 40~pc of the Sun as part of the NStars program. These nearby stars have good-quality parallaxes ($\sigma_\pi/\pi~\lesssim~0.05$), and are all found in the 2MASS catalog. Uncertainties on the stellar parameters for individual stars are not provided in the catalog; we choose to set them to $\sigma_{\rm Teff} = 100$~K, $\sigma_{\rm log~g} = 0.3$ dex, and $\sigma_{\rm [Fe/H]} = 0.2$ dex. Furthermore, we cannot rely on the method used for LAMOST data, where we selected stars from the same observed plate to derive the selection and luminosity functions. For this and all subsequent tests of our code, we obtain the underlying luminosity distribution using only the color-magnitude selection from the algorithm outlined in Section~2.3, which we apply to all stars in the test catalog without regard to position on the sky. Figure~\ref{hiptest} shows the results of running our distance code using the Gray et al. stellar parameters plus 2MASS magnitudes and colors for stars with temperature in the range $3500$~K$ < T_{\rm eff} < 9000$~K. This temperature cut reduces the sample from a total of 1525 stars to 1199 used for Figure~\ref{hiptest}. The histogram compares the known distances, $d_{\rm Hipparcos}$ (from the trigonometric parallaxes) to the derived distances, $d_{\rm LAMOST}$, expressed as a residual: $(d_{\rm LAMOST} - d_{\rm Hipparcos})/d_{\rm Hipparcos}$.
The residuals are centered on zero (i.e., no systematic offset is present), with a scatter of $\sim17\%$ (median absolute deviation of the data; we use this instead of fitting a Gaussian because the residuals are obviously asymmetric and non-Gaussian). Because of their location in the Solar neighborhood, the majority of these stars are metal-rich main-sequence stars (1151 are dwarfs with $\log{g}<3.5$, and 48 are giants). Thus, while these stars are a useful test, they do not explore the variety of stellar populations we expect to find in a survey such as LAMOST.

To examine our algorithm's behavior on a more heterogeneous data set, we test our code on two samples of RAVE DR2 stars. The first sample is the RAVE-6D catalog: \url{http://www.astro.rug.nl/~rave/}, which is from \citet{bsh+10}. We use the RAVE stellar parameters from this table as inputs to the distance code. Uncertainties on the stellar parameters were set to $\sigma_{\rm Teff} = 100$~K, $\sigma_{\rm log~g} = 0.3$~dex, and $\sigma_{\rm [Fe/H]} = 0.2$~dex to approximate the typical errors in RAVE.  The left panel of Figure~\ref{ravetest} shows residuals from a comparison of our results to those from \citet{bsh+10}. There is a systematic shift of $\sim26\%$ between our distances and those of \citet{bsh+10}, with scatter of $\sim23\%$. 
On close examination, there is no obvious correlation of the distance residuals with any of the input stellar parameters (e.g., $\log{g}$, [Fe/H]). Thus the systematic offset between our distance scale and that of \citet{bsh+10} may be due to differences in the isochrones used in the fitting. \citet{bsh+10} used Yale-Yonsei isochrones in a grid with 40 logarithmically spaced ages between 0.01 and 15~Gyr. The differences between the Dartmouth and Yale-Yonsei isochrones, and the heavy emphasis on very young ages in the \citet{bsh+10} grid, seem to cause systematic shifts. Reassuringly, when we applied our $\chi^2$ code, which defines $\chi^2$ in the same way as \citet{bsh+10}, to these data, the scatter about the mean difference is small (but the systematic offset remains).

The second set of RAVE data on which we tested the code is the catalog of \citet{zmb+10}. These authors improved upon the method of \citet{bsh+10}
 by using a linearly spaced grid of ages, deriving distances separately using Yale-Yonsei, Padova, and Dartmouth isochrones, and by weighting stages of stellar evolution to account for the relative numbers of stars of different masses. We ran our code on the parameters of roughly 16,000 stars ($\sim5800$ giants and $\sim9900$ dwarfs) from \citet{zmb+10} and compared directly to their Dartmouth results. The comparison is shown in the right panel of Figure~\ref{ravetest}, with residuals calculated in the same way as for the Breddels et al. sample.
Again, these residuals show a systematic shift such that our estimates are lower than the RAVE distances. The systematic difference between our result and the RAVE distances is smaller than for the comparison to \citet{bsh+10}, as expected since the grid spacing in age is similar and we are using the same isochrone sets (Dartmouth).  Our Bayesian method produces decidedly non-Gaussian residuals, with a median offset of $\sim12\%$ and scatter of $\sim16\%$.

As a final test of the accuracy of our derived distances, we generate catalogs using the Besan\c{c}on model of the Milky Way \citep{rrd+03} for two fields of view: $(l, b) = (180^\circ, 30^\circ)$ and $(l, b) = (180^\circ, 60^\circ)$. The $b=60^\circ$ catalog contains 5614 stars (299 giants and 5315 dwarfs) with $3500$~K$ < T_{\rm eff} < 10000$~K; the $b=30^\circ$ field has 469 giants and 13384 dwarfs in the same temperature range. We again assign uncertainties to the stellar parameters of $\sigma_{\rm Teff} = 100$~K, $\sigma_{\rm log~g} = 0.3$~dex, and $\sigma_{\rm [Fe/H]} = 0.2$~dex. Figure~\ref{disttest} shows the results of running our algorithm with the stellar parameters from the Besan\c{c}on model as inputs. We recover the distances well, 
with a mildly bimodal distribution and a slight tail extending to $\sim50\%$ overestimate of distances. These residuals have roughly zero median offset, but it is clear that a large fraction of stars have distances underestimated by $\sim5\%$. The positive ``bump'' in these residuals appears to consist mostly of the youngest nearby stars and old, metal-poor halo giants in the Besan\c{c}on model catalog.

\section{Application to LAMOST data}\label{sec:lamost_dist}

Having verified the effectiveness of our distance code on catalog and simulated data, we now apply the code to the existing LAMOST data. As of this date, the LAMOST catalog (internal data releases 1 and 2) consists of $\sim1.8$~million stars with stellar parameters (out of $\sim3.6$~million that have been observed; stars with low S/N, cool M-type stars, and hot OBA stars do not have parameters from the LAMOST pipeline). In this section, we show some simple ``sanity checks'' to verify that the code is producing reasonable results, and to provide an idea of the scope of the LAMOST data set.

\begin{figure}[!t]
\begin{center}
\includegraphics[width=0.9\columnwidth]{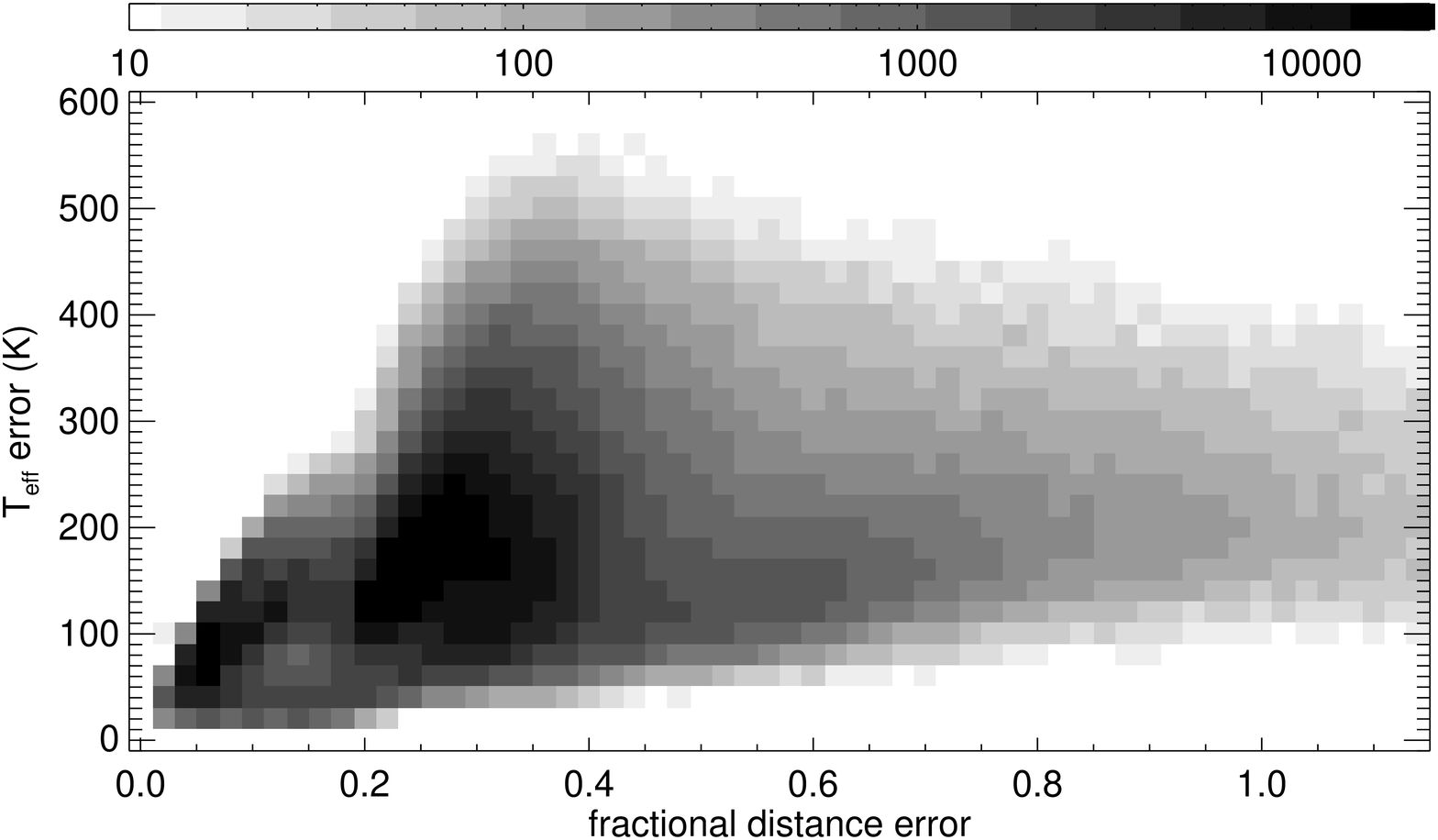} \\
\includegraphics[width=0.9\columnwidth]{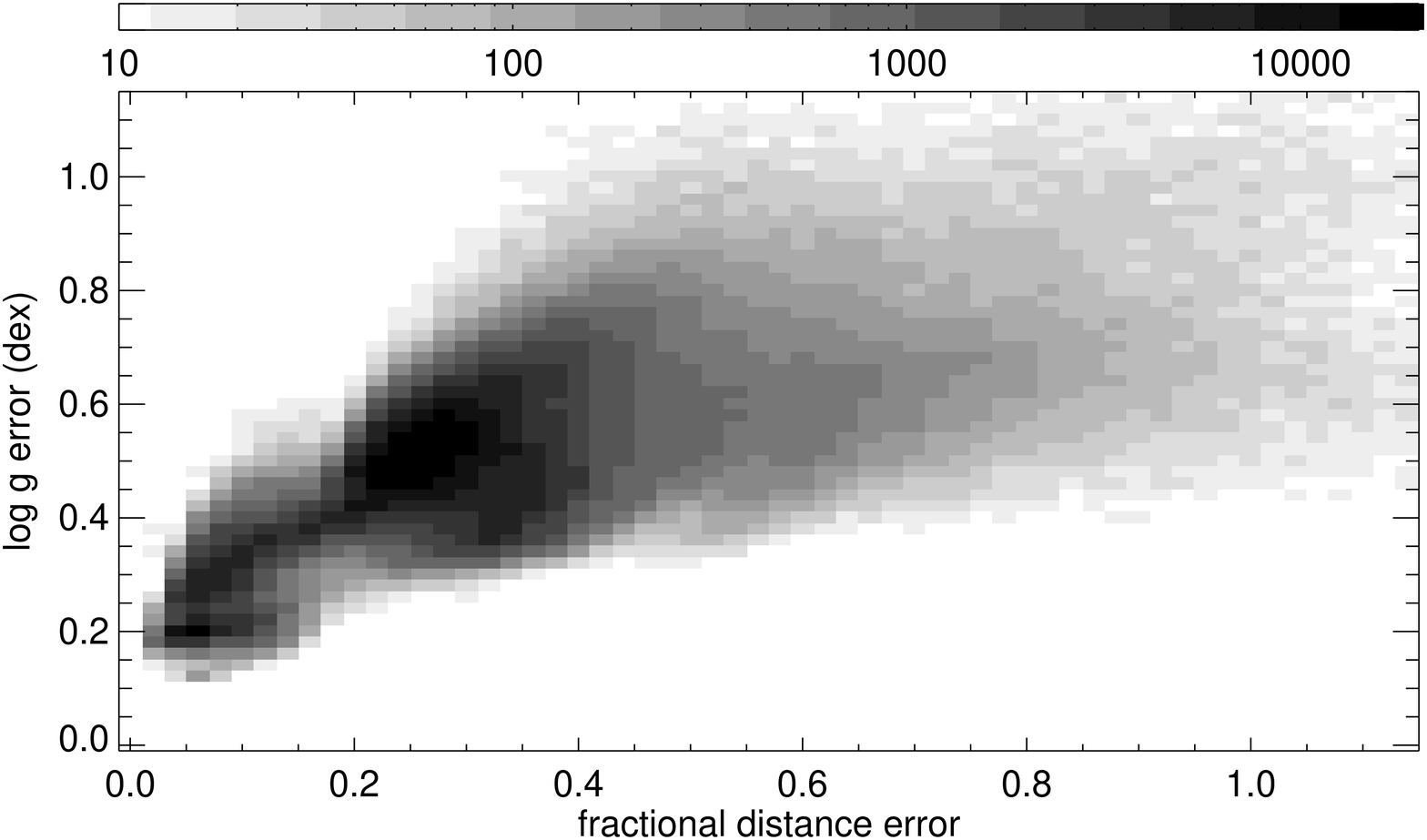} \\
\includegraphics[width=0.9\columnwidth]{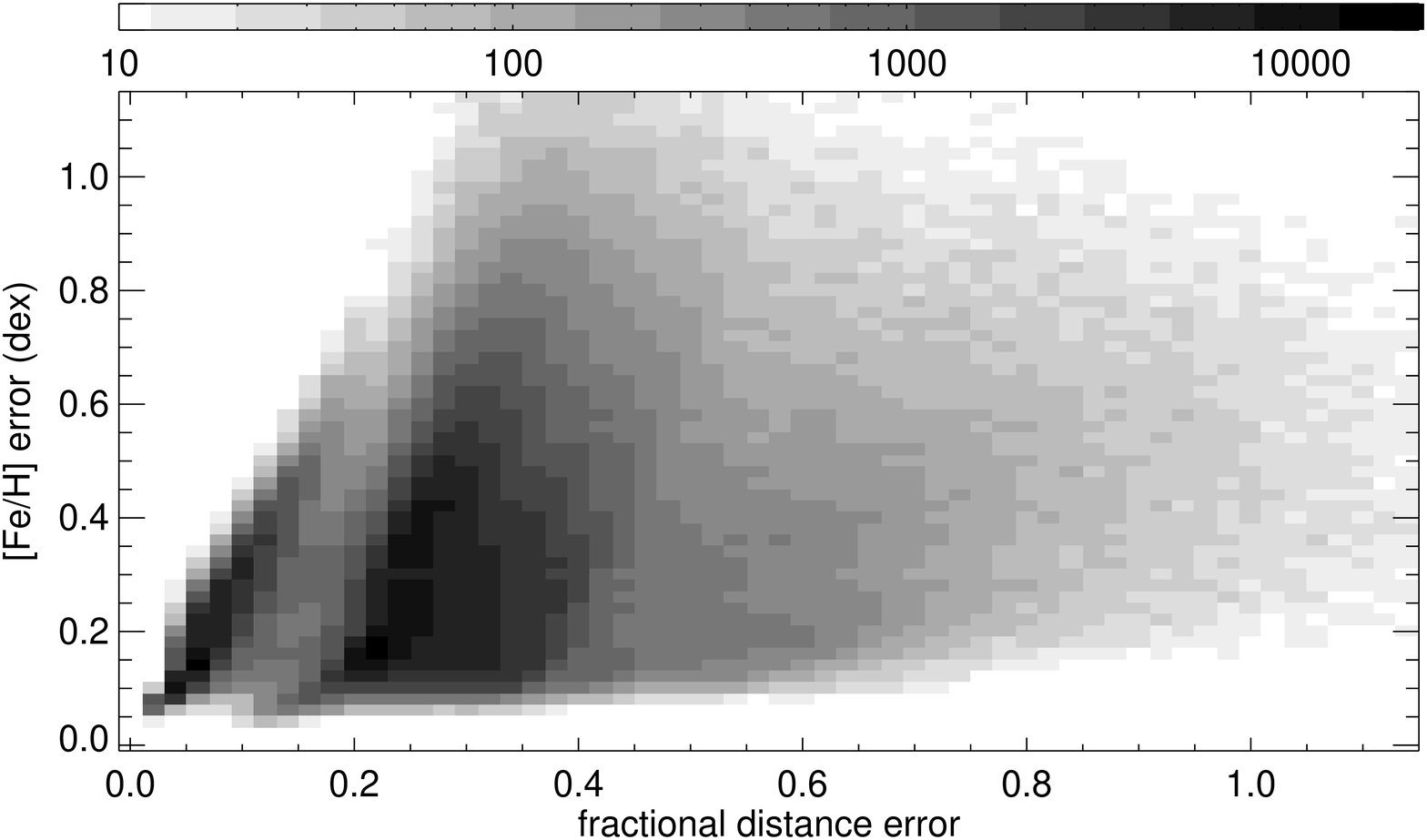}
\caption{Fractional errors on the derived distances for LAMOST stars compared to the uncertainties on the stellar parameters $T_{\rm eff}$, $\log{g}$, and [Fe/H] (from top to bottom). The grayscale encodes the number of stars in each bin on a logarithmic scale between 10 and 20,000. The distance errors are only weakly correlated with $T_{\rm eff}$ or [Fe/H] uncertainties. By far the strongest dependence is seen in the center panel, which shows a roughly linear correlation between surface gravity uncertainty and the error in the derived distance.
}
\label{param_dist_err}
\end{center}
\end{figure}

\subsection{LAMOST Stellar Parameters}

The LAMOST parameters for stars in the range $3500$~K$ < T_{\rm eff} < 8000$~K and with S/N$ > 5$ in $g$ and $r$ bands have median uncertainties of  $\sim160$~K, $\sim0.5$~dex, and $\sim0.3$~dex, in $T_{\rm eff}$, $\log{g}$, and [Fe/H], respectively. We note that the [Fe/H] uncertainties in the second full year of survey operations (2013 September -- 2014 June) are significantly smaller (median $0.18$~dex) than the earlier periods; the $T_{\rm eff}$ and $\log{g}$ errors are similar in earlier and later data. It is unclear whether this is due to changes in the LAMOST data reduction pipeline, or improved data quality as the survey progresses.

The parameter that most strongly affects the derived distance errors is surface gravity. This can be seen in Figure~\ref{param_dist_err}, which compares the errors on $T_{\rm eff}$, $\log{g}$, and [Fe/H] with the error in derived distances based on those parameters. There is a slight correlation of $\sigma_{\rm Teff}$ and $\sigma_{\rm d}$, but little dependence of distance errors on uncertainty in [Fe/H]. The middle panel, showing $\sigma_{\rm log~g}$ vs. $\sigma_{\rm d}$, exhibits a roughly linear correlation between the surface gravity uncertainty and the errors on the derived distances. For an uncertainty of $\sim0.5$~dex in $\log{g}$, Figure~\ref{param_dist_err} suggests that we can expect a $\sim25-35\%$ distance error. It is thus vital that surface gravities from LAMOST spectra are determined as precisely as possible. \citet{lfw+14} recently published a method to improve $\log{g}$ estimates for giant stars in the Kepler field that have also been observed with LAMOST. Based on corrections from comparison to asteroseismic $\log{g}$ measurements from Kepler, Liu~et~al. obtain uncertainties in $\log{g}$ from LAMOST spectra of $\sim0.1$~dex, which yields distance estimates with better than 10\% precision. Indeed, at a given temperature, $\Delta M_{\rm abs} \sim 2.5 \Delta (\log{g}),$\footnote{Because $M_{\rm abs}\propto -2.5 \log{L}$, $L \propto R^2$, and $g \propto R^{-2}$. Thus $M_{\rm abs} \propto 2.5 \log{g}$.} so if the uncertainty of $\log{g}$ is improved by 0.1 dex, the uncertainty in absolute magnitude improves by 0.25 mag, and the accuracy of the distance estimate improves by $\sim$12\%.  

\begin{figure}[!t]
\begin{center}
\includegraphics[width=0.95\columnwidth]{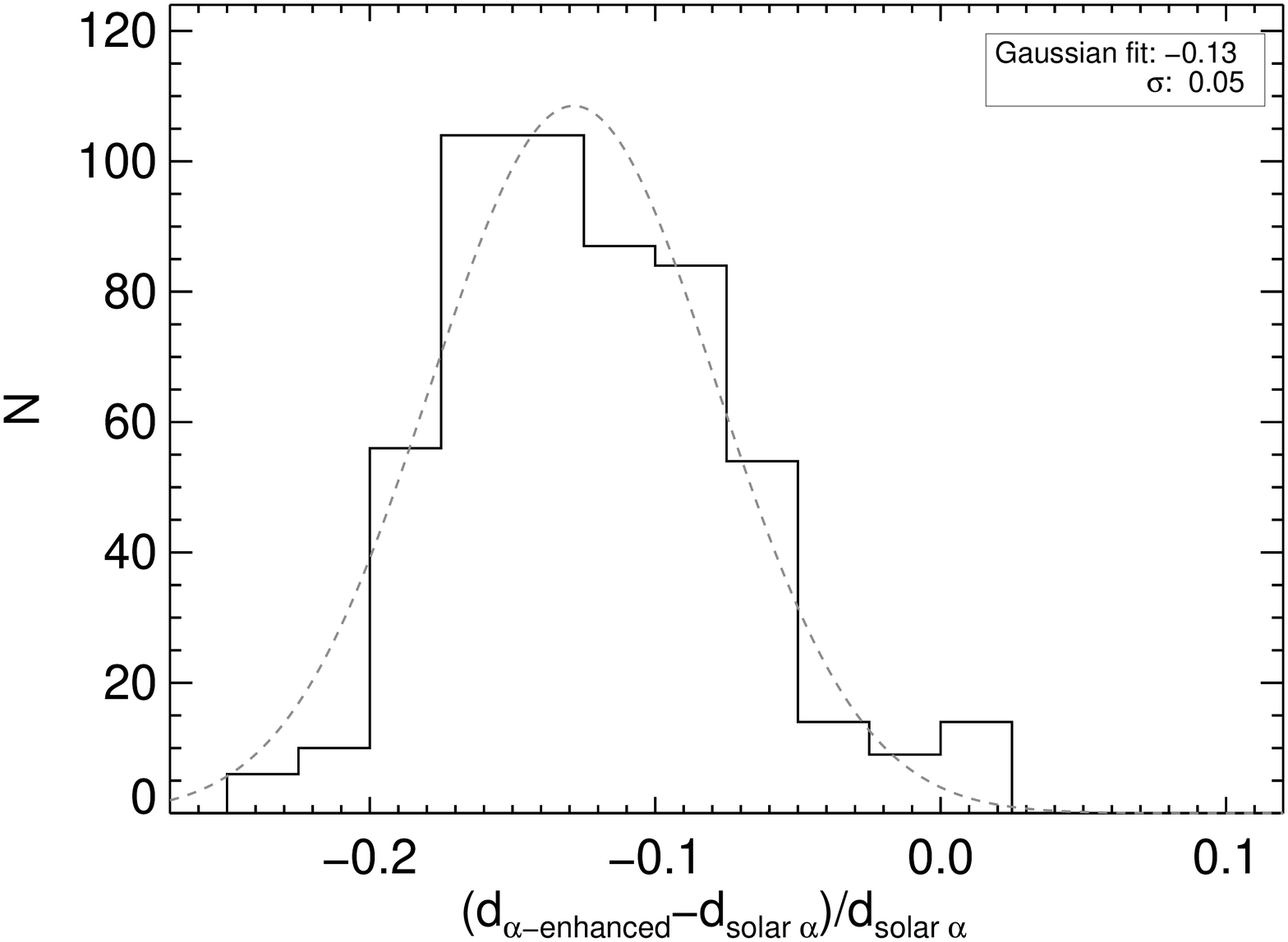}
\caption{Difference between distances to halo giants ($\left|Z\right| > 3$~kpc, $-2.4 < $[Fe/H]$ < -1.0$, with S/N$ > 10$ in $g$ and $r$-band) measured with solar-scaled isochrones ($[\alpha$/Fe] = 0.0; labeled ``solar $\alpha$'') and an $\alpha$-enhanced ($[\alpha$/Fe] = 0.4) grid of isochrones. On average, the $\alpha$-enhanced isochrones find distances $\sim13\%$ smaller (derived via the dashed-line Gaussian fit shown above) than those from the solar-scaled grid.
}
\label{alpha_enhanced_dist}
\end{center}
\end{figure}

\subsection{Effect of $\alpha$-element Abundances on Distances to Metal-poor Halo Giants}

As noted in Section~2, our algorithm tends to overestimate the distances to metal-poor halo giants in synthetic catalogs from the Besan\c{c}on model. It is well established that the metal-poor stellar populations of the Milky Way halo are typically enhanced in $\alpha$-elements relative to disk populations (e.g., \citealt{vis+04}), with metal-poor ([Fe/H]$ < -1.0$) halo stars typically having $[\alpha$/Fe]$\approx0.4$. We now return to a subset of stars for which we have LAMOST stellar parameters, and examine the effect of replacing the solar-scaled isochrones with $\alpha$-enhanced versions in our distance code. To do so, we generate a new isochrone grid with $[\alpha$/Fe] = +0.4, [Fe/H]$ < -1.0$, and the same steps in age as the original isochrone set. We run our distance algorithm on a set of stellar parameters from 239,446 LAMOST spectra (comprised of recent, third-year LAMOST spectra) with the $\alpha$-enhanced isochrones. From the resulting distance catalog, we then select out only likely metal-poor halo stars with S/N$ > 10$ in $g, r$-bands, $-2.4 < $[Fe/H]$ < -1.0$, $\log{g} < 3.5$, and at least 3~kpc from the Galactic plane. This produces a sample of 542 likely halo stars. Figure~\ref{alpha_enhanced_dist} compares the distance from the $\alpha$-enhanced grid to the distance from the original isochrone grid, in the sense $(d_{\rm \alpha-enhanced} - d_{\rm solar~\alpha})/d_{\rm solar~\alpha}$. We find that the $[\alpha$/Fe] = +0.4 grid produces distances on average 13\% nearer than those from the $[\alpha$/Fe] = 0.0 grid. This likely explains the $\sim20\%$ systematic overestimation of distances for halo stars from the Besan\c{c}on catalogs. Because halo populations in the Besan\c{c}on model were oxygen-enhanced relative to disk populations \citep{rrd+03}, our assumption of solar $\alpha$-abundances likely biases the derived distances. Adopting a more appropriate $\alpha$-enhanced isochrone grid for metal-poor halo stars would remedy this situation. Indeed, one would ideally incorporate a measured $[\alpha$/Fe] from the LAMOST spectrum itself into the distance estimation for each star; we will include this in future upgrades to the distance code as abundance estimates become available for LAMOST stars.

\begin{figure}[!t]
\begin{center}
\includegraphics[width=0.8\columnwidth]{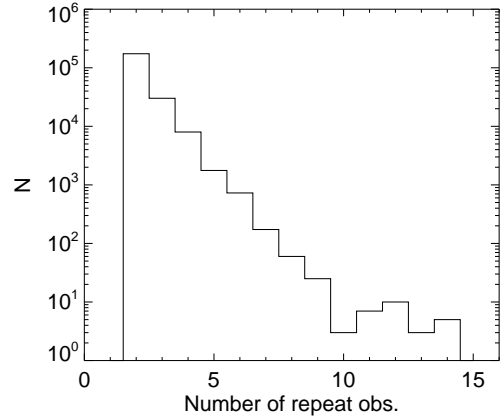}
\caption{Number of repeat observations of the 214,514 unique stars with multiple spectra in the LAMOST database. The majority of these objects have fewer than 4 observations, but some have as many as 14 separate spectra.
}
\label{lamost_ngrps}
\end{center}
\end{figure}

\subsection{Internal LAMOST Checks on Repeat Observations}

\begin{figure}[!t]
\begin{center}
\includegraphics[width=1.0\columnwidth]{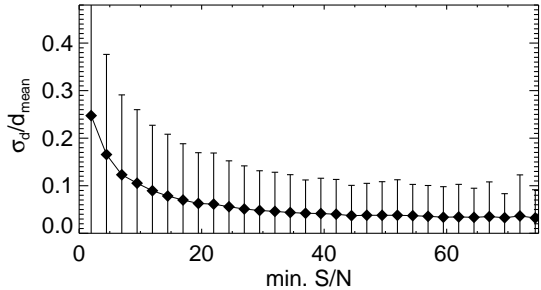}
\caption{Standard deviation $\sigma_d$ of the distance estimate for the 214,514 stars with multiple measurements in LAMOST. This is expressed as a fractional deviation of the distance, $\sigma_d / d_{\rm mean}$, as a function of the minimum signal to noise of the spectra included in the derivation of $\sigma_d$ for each star. We calculated mean scatter (filled diamonds) and its standard deviation (error bars) for these results in bins of 2.5 in S/N. The scatter from repeat measurements is typically $\sim5\%$ for high $S/N$ stars (min. S/N$ \gtrsim 20$), then increases to $\sim20\%$ at the low S/N end. This suggests that even for fairly poor quality spectra, our distance derivations (and thus the stellar parameters on which they are based) are robustly repeatable. 
}
\label{lamostrepeats}
\end{center}
\end{figure}

Of the $\sim1.8$~million stellar spectra in the LAMOST catalog, $\sim550,000$ of them ($\sim30\%$) are stars with repeat observations. There are 214,514 unique stars that have been observed multiple times and have sufficient quality spectra at each epoch to derive stellar parameters. An individual star may have as many as 14 observations, but most have 2-4 observations; the distribution of the number of repeat measurements is shown in Figure~\ref{lamost_ngrps}. Figure~\ref{lamostrepeats} shows the standard deviation of our distance measurements for stars with multiple observations. This is expressed as a fractional deviation of the mean measured distance, $\sigma_{\rm d}/$d$_{\rm mean}$, and plotted as a function of the minimum signal to noise of the measurements being compared. One would expect that the scatter in derived distances would increase if one (or more) of the spectra has low S/N. This is precisely what is seen in Figure~\ref{lamostrepeats} -- the scatter is $\sim5\%$ for spectra with minimum S/N$>20$, and begins to rise for S/N below 20. However, even when the minimum S/N is as low as 2.5, the typical scatter in distances in only $\sim20\%$. This verifies that (a) our code produces repeatable results when applied to multiple observations of the same star, and (b) the LAMOST pipeline provides consistent estimates of stellar parameters from these multiple observations.

\begin{figure}[!t]
\begin{center}
\includegraphics[width=1.0\columnwidth]{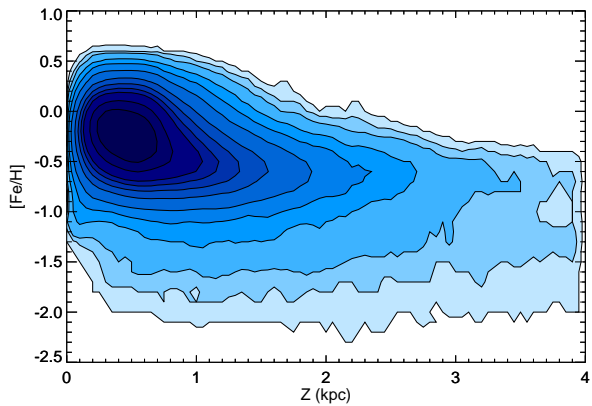}
\caption{Metallicity from the LAMOST pipeline versus height above the plane for a sample of 189,106 stars at $b > 60^\circ$ that have S/N$_g>10$. The $Z$ coordinate is based on distances derived by our code using LAMOST stellar parameters. Contours contain (2, 5, 10, 25, 50, 100, 200, 300, 400, 500, 750, 1000, 1500, 2500, 4000) stars. As expected for disk stars, the mean metallicity falls from near solar just above the Galactic plane to $\langle{\rm [Fe/H]}\rangle  \approx -0.6$ near $Z \sim 1$~kpc. This peak metallicity, which is typical of the Galactic thick disk, persists as far out as we probe, with a long tail to lower metallicity. }
\label{feh_vs_z_hilat}
\end{center}
\end{figure}

\subsection{Results from LAMOST Data}

After running our distance code on the entire catalog of LAMOST stellar parameters, we perform some checks to verify that the results make sense, and to explore the utility of our distances for Galactic structure studies. Using our distances, we calculate Galactocentric Cartesian coordinates (assuming the Sun is at $R_0 = 8$~kpc, with $(X, Y, Z)_{\rm Sun} = (-8, 0, 0)$~kpc). The first test is to see whether the metallicity distribution as a function of height above the Galactic plane near the north Galactic cap is as expected. We select stars at $b > 60^\circ$, keeping only those with S/N$ > 10$ in the SDSS $g$-band. This yields 189,106 stars. This sample should roughly probe the Galactic metallicity gradient with height; one expects that on average the metallicity should be nearly solar close to the plane, and decrease with height as the thin disk transitions into the lower-metallicity thick disk. Indeed, this is exactly what is seen in a contour plot of these data in Figure~\ref{feh_vs_z_hilat}. The peak metallicity decreases from slightly subsolar at $Z \sim 0.3$~kpc to [Fe/H]$ \sim -0.6$ at $Z \sim 1$~kpc. Above this, the peak metallicity remains roughly the same, with a long tail to low metallicities representing predominantly local halo stars.

\begin{figure}[!t]
\includegraphics[width=1.0\columnwidth]{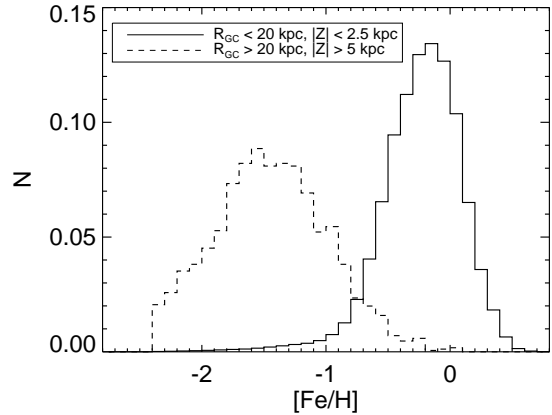}
\caption{Normalized metallicity distribution of 1,473,135 stars at $R_{\rm GC} < 20$~kpc and $\left|Z\right| > 2.5$~kpc (solid histogram). The 1,705 stars at $\left|Z\right| > 5$~kpc and $R_{\rm GC} > 20$~kpc are represented by the dashed line. The latter should be mostly halo stars, and peaks at [Fe/H]$ \sim -1.5$ as expected for the Galactic halo, with few metal-rich ([Fe/H]$~>~-1.0$) stars. The $\left|Z\right| < 2.5$~kpc sample contains mostly metal-rich stars, as expected for predominantly disk populations. The abrupt cutoff at [Fe/H]$ = -2.5$ is due to the lower limit of metallicities produced by the LAMOST pipeline rather than a real effect. \citet{lbc+15} adapted the Sloan SSPP for more general usage; use of this pipeline on LAMOST spectra avoids the artificial cutoff at [Fe/H]$ = -2.5$. }
\label{hiz_feh}
\end{figure}

Though giant stars in the Galactic halo represent a tiny fraction of the stars observed by LAMOST, we also hope to use them to explore structure (and substructure) in the halo. We thus wish to check whether our distances can be used to isolate a relatively pure sample of Milky Way halo giants. To test this, we 
select stars with Galactocentric radii $R_{\rm GC} > 20$~kpc that are also at heights $\left | Z_{\rm GC} \right | > 5$~kpc above/below the plane. Such a sample of stars should be predominantly halo stars. We check this by plotting a metallicity histogram (dashed line in Figure~\ref{hiz_feh}) for the 1528 stars selected in this way. These stars peak at a metallicity around [Fe/H]~$\sim -1.5$, as expected for inner-halo stars, with very few metal-rich stars. In contrast, a sample selected to be inside $R_{\rm GC} < 20$~kpc and near the disk ($\left | Z_{\rm GC} \right | < 2.5$~kpc; solid line in Figure~\ref{hiz_feh}) contains mostly metal-rich stars with disk-like [Fe/H].

Note that neither of these sanity checks showing metallicity distributions for different Galactic populations (Figures~\ref{feh_vs_z_hilat} and \ref{hiz_feh}) represents the true Galactic metallicity distribution for these populations. To derive the intrinsic distribution would require correcting the selection effects present in LAMOST data. These figures are simply meant to illustrate that stellar samples selected using our derived distances have properties similar to what one might expect based on our knowledge of metallicity distributions of Milky Way components.

\begin{figure}[!t]
\begin{center}
\includegraphics[width=1.0\columnwidth]{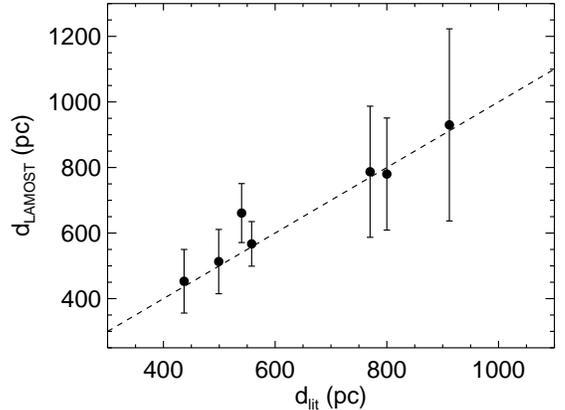}
\caption{Comparison of our derived distances ($d_{\rm LAMOST}$) to those from the literature \citep[$d_{\rm lit}$]{dam+02} for seven open clusters found in LAMOST. Points represent the central value of the best-fitting Gaussian for each cluster's distance distribution, and error bars show the Gaussian $\sigma$. The dashed line represents one-to-one agreement. All but one of the clusters agrees very closely with the known distance.
}
\label{lamost_ocs}
\end{center}
\end{figure}

Finally, we search the LAMOST database for open cluster member stars. We begin with the compilation of known Galactic open clusters available at \url{http://www.astro.iag.usp.br/ocdb/} \citep{dam+02}. For each cluster in this list, we initially selected all stars from LAMOST within the published cluster diameter that also have LAMOST RV within 20 km~s$^{-1}$ of the published value (note that we only used clusters with known RVs for this exercise). After this initial cut, we examined histograms of velocities, distances, and metallicities for each of the clusters with more than 15 candidates. For clusters with obvious distance and velocity peaks, we manually select stars within $\sim10$~km~s$^{-1}$ of the peak value, and fit a Gaussian to the distance distribution of these cluster candidates. Clusters with obvious signatures were NGC~1039, NGC~1662, NGC~2168, NGC~2281, NGC~2548, ASCC~26, and NGC~1647. The number of candidates selected ranged from 19 to 102 stars, with the nearest clusters having the most candidates. Figure~\ref{lamost_ocs} compares our measured distances (from the Gaussian fits) for these seven clusters, $d_{\rm LAMOST}$, to those from \citet{dam+02}, $d_{\rm lit}$. Error bars on these points represent the Gaussian $\sigma$ of the stars included. The dashed line corresponds to one-to-one agreement between our measurements and literature values. All but one of the seven clusters' distances is consistent with values from the literature. Thus, we have confirmed the effectiveness of our distance estimations. This simple exercise highlights the potential of LAMOST to amass a sample of open cluster stars with homogeneously measured metallicities, velocities, and distances that can be used to probe the Galactic disk in exquisite detail. 

\section{Conclusions}

We present a method to derive distances to stars with measured stellar parameters ($T_{\rm eff}$, $\log {g}$, and [Fe/H]). This was developed with particular interest in deriving distances to the many millions of stars that will be observed by the LAMOST survey, in order to enable studies of Galactic structure with this vast data set. The code is based on a Bayesian method that evaluates the posterior PDF in absolute magnitude for a given star, estimated via comparison to a grid of theoretical isochrones. The PDF incorporates information about $T_{\rm eff}$, $\log {g}$, and [Fe/H], along with their uncertainties. To account for selection effects, we take advantage of the fact that each LAMOST plate typically observes $\sim3000$ stars simultaneously. The observed distribution of $\log{g}$ for stars within 0.25 magnitudes in color-magnitude space is mapped onto theoretical isochrones to derive a proxy ``luminosity function'' expected for stars in that region of sky at the given color and magnitude. This accounts simultaneously for the selection function through which stars were chosen for LAMOST observation, and the variation in stellar populations with Galactic line of sight (and distance). 
A flat age prior is implemented, since we have no information about the ages of individual stars. This could, in principle, be modified to account for the relatively well-known age-metallicity relation in the Milky Way, but we choose to leave it flat so as not to bias studies of Galactic stellar populations based on LAMOST data. Likewise, we do not impose any priors related to Milky Way stellar populations; since we wish to study Galactic structure, we prefer to avoid introducing assumptions into our distance calculations.

We test our code by measuring distances to samples of stars from Hipparcos and RAVE that have known stellar parameters. Our distances agree with the parallax distances from Hipparcos, with roughly 17\% scatter in the residuals. We find a 12\% systematic shift between our distances and those measured by \citet{zmb+10} from the same RAVE sample, with only 16\% scatter, but with a large tail toward underestimate of the distance. We also test our code on simulated data along two lines of sight from the Besan\c{c}on model, and find that we recover the model distances with no net offset. There is, however, an apparent $\sim20\%$ overestimate of distances to distant, metal-poor halo stars by our code. The source of this systematic shift is unclear, but it may be due to the fact that we have used isochrones with solar $\alpha$-element abundances. Stars in the Milky Way halo are, on average, old and $\alpha$-enhanced relative to the Solar neighborhood (see, e.g., \citealt{vis+04}). Because the RGB of an isochrone is shifted slightly to fainter absolute magnitudes when $[\alpha$/Fe] is increased (at fixed [Fe/H] and age), the use of Solar-scaled isochrones will tend to bias results for $\alpha$-rich stars toward overestimation of distances. We confirmed this impression by re-running the distance derivation for a subset of metal-poor halo giants, using an $\alpha$-enhanced isochrone grid ($[\alpha$/Fe]$=+0.4$). Indeed, this exercise shifted the distance estimates for these stars by an average of 13\% closer than the estimates based on the Solar-alpha grid. Thus we suggest that adopting a more appropriate $\alpha$-enhanced isochrone set for metal-poor halo stars would remedy the overestimation of distances to these stars, and that ultimately the measured [$\alpha$/Fe] should be incorporated into the distance estimation process.

Finally, we present some results based on LAMOST data. A sample of $\sim189,000$ stars near the north Galactic cap shows expected trends in [Fe/H] with height ($Z$) above the Galactic plane. We also show that a sample selected to be distant halo stars based on our derived distances consists solely of metal-poor stars with a distribution peaked around [Fe/H]$ \sim -1.5$, as expected for the Galactic inner halo. Some studies of kinematics of nearby stars based on LAMOST data and with distances from this code have already appeared in the literature \citep{tlc+15,xia15}, and we anticipate that distances derived by this method will prove useful for numerous upcoming studies of Galactic structure. Furthermore, as the LAMOST data reduction pipelines continue to improve, we anticipate that uncertainties on stellar parameters derived from the spectra will become smaller, which will in turn improve our estimates of the distances. Eventually, parallaxes that will come from the $Gaia$ mission \citep{perryman2001} will likely supersede these distance measurements for the majority of the stars observed by LAMOST, as part of a vast sample of direct distance measurements throughout the Galaxy.

\acknowledgements

We thank the anonymous referee for careful and thoughtful comments. This work was supported by the U.S. National Science Foundation under grants AST 09-37523 and AST 14-09421. C.~L. also acknowledges the Strategic Priority Research Program ``The Emergence of Cosmological Structures'' of the Chinese Academy of Sciences, grant No. XDB09000000, the National Key Basic Research Program of China, grants No. 2014CB845700, and the National Science Foundation of China, grants No. 11373032 and 11333003. T.~C.~B. acknowledges partial support from grant PHY 08-22648: Physics Frontiers Center/Joint Institute for Nuclear Astrophysics (JINA), and PHY 14-30152; Physics Frontier Center/JINA Center for the Evolution of the Elements (JINA-CEE), awarded by the U.S. National Science
Foundation. W.~Y. appreciates support from the National Science Foundation of China, grant No. 11403056. Guoshoujing Telescope (the Large Sky Area Multi-Object Fiber Spectroscopic Telescope LAMOST) is a National Major Scientific Project built by the Chinese Academy of Sciences. Funding for the project has been provided by the National Development and Reform Commission. LAMOST is operated and managed by the National Astronomical Observatories, Chinese Academy of Sciences. This publication makes use of data products from the Two Micron All Sky Survey, which is a joint project of the University of Massachusetts and the Infrared Processing and Analysis Center/California Institute of Technology, funded by the National Aeronautics and Space Administration and the National Science Foundation.

\bibliographystyle{apj}

\end{document}